\documentclass{actasen}

\begin{document}

\setcounter{page}{000}

\Volume{2014}{000}

\runheading{N. Yasutake et al.}%

\title{Quark-Hadron Phase Transition \\ with Finite-Size Effects in Neutron Stars$
^{\dag}~ 
$}

\footnotetext{$^{\dag}$ This work was supported by JSPS KAKENHI Grant Numbers 25105510, 23540325, 24105008.


\hspace*{5mm}$^{\bigtriangleup}$ nobutoshi.yasutake@p.chibakoudai.jp\\

\noindent 0275-1062/01/\$-see front matter $\copyright$ 2011 Elsevier
Science B. V. All rights reserved. 

\noindent PII: }

\enauthor{N.~Yasutake$^{\bigtriangleup, 1}$, S.~Benic$^2$, D.~Blaschke$^{3,4}$, T.~Maruyama$^5$, T.~Tatsumi$^6$}
{1.~Physics Department, Chiba Institute of Technology, \\ Shibazono 2-1-1, Narashino, Chiba, 275-0023, Japan\\
2.~Physics Department, Faculty of Science, University of Zagreb, Zagreb 10000, Croatia\\
3.~Institute of Theoretical Physics, University of Wroc{\l}aw, 50-204 Wroc{\l}aw, Poland\\
4.~Bogoliubov Laboratory of Theoretical Physics, JINR Dubna, 141980 Dubna, Russia\\
5.~Advanced Science Research Center, Japan Atomic Energy Agency, \\Tokai, Ibaraki 319-1195, Japan\\
6.~Department of Physics, Kyoto University, Kyoto 606-8502, Japan}

\abstract{We study the quark-hadron phase transition with the finite-size effects in neutron stars. 
The finite-size effects should be, generally, taken into account in the phase transition of multi-component system. 
The behavior of the phase transition, however, strongly depends on the models for quark 
and hadron matter, surface tension, neutrino fraction, and temperature. 
We find that, if the surface tension is strong, the EOS becomes similar to the case of a 
Maxwell construction for any hadron and/or quark model, though we adopt the Gibbs conditions.
We also find that the mass-radius relations for that EOS are consistent with the observations, 
and our  model is then applicable to realistic astrophysical phenomena such as the thermal 
evolution of compact stars.}


\maketitle

\section{INTRODUCTION}
The equation of state~(EOS) is one of the most important topics in studies on neutron stars. 
But there is a large uncertainty in the finite-density region of the EOS. 
One way of advancing our understanding of the EOS of neutron stars~(NSs) is to study 
baryon-baryon(BB) interactions by Lattice QCD~(LQCD) simulations~\cite{1}, another to perform 
experiments such as heavy-ion collisions at JPARC. 
Since the quantum-many body methods, such as variational principle~\cite{2,3,4}, 
Brueckner-Hartree-Fock(BHF) theory~\cite{5,6,7}
and Dirac-Brueckner-Hartree-Fock theory~\cite{8}, which reveal the EOS for NSs using BB 
interaction directly are applicable only to hadron matter, there is the need for developing 
methods to investigate the possible the existence of exotic matter such as quark matter 
in NSs \cite{8a}.

In this study, we consider nucleon and quark degrees of freedom, and study the mixed phase 
of the quark-hadron phase transition in NSs. The details are given in Refs.~\cite{9a,9b,9}.

\section{OUR MODELS and NUMERICAL RESULTS}

We adopt the BHF theory for hadronic matter~\cite{5,6} with the Bonn B potential~\cite{10}. 
New potentials could be employed when LQCD simulations and/or experiments would reveal them. 
For the quark matter we adopt a non-local NJL model~\cite{11,12} with vector coupling. 
In Fig.~1 we show results using for the ratio of the 
vector and the scalar channel couplings $\eta_V = G_V /G_S = 0.20$. 

Note that the phase transition itself has large uncertainty~\cite{13} which comes mainly 
from {\it finite size effects}, where inhomogeneous structures appear depending on the 
balance between surface tension and Coulomb interaction \cite{9a,14}. 
We assume a sharp boundary between quark and hadron matter, with values for the surface tension
parameter $\sigma$=10 MeV fm$^{-2}$ and  $\sigma$=40 MeV fm$^{-2}$, and discuss the effects 
of its variation as in our previous studies with a simple quark model~[15,16]. 

The numerical results for the EOS with quark-hadron phase transition are shown in Fig.~1. 
In this figure, it is shown that the strong surface tension ($\sigma$=40 MeV fm$^{-2}$) 
makes the phase transition similar to the Maxwell construction one. 
This behaviour is general since it is also found in the simple bag model~\cite{9a,15,16}.
We obtained mass-radius relations for our EOS with maximum masses $M_{{max}}$ consistent 
with the observational results~\cite{17,18}, i.e., $M_{{max}}=2.47~(2.49)~M_\odot$ for 
$\sigma=10~(40)~$ MeV fm$^{-2}$ and $\eta_V=0.20$. 
The radii of NSs in our models are 12-14 km.

\begin{figure}[tbph]
\centering
{\includegraphics[angle=0,width=6cm]{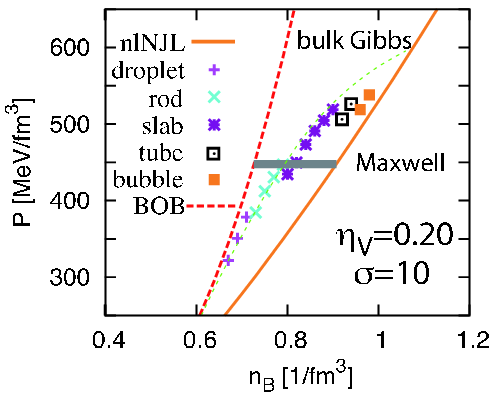}}~~~~
{\includegraphics[angle=0,width=6cm]{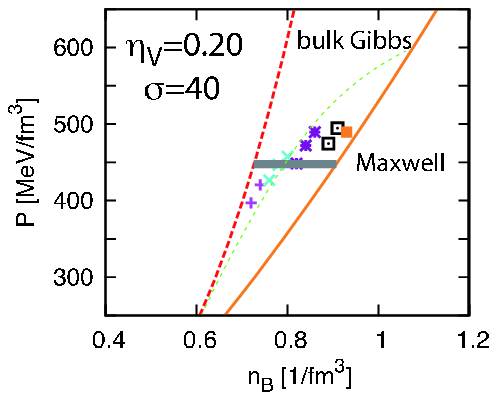}}
\vspace{-2mm}
\caption{EOS with quark-hadron phase transition shown as pressure vs. baryon number density
with pure quark matter from non-local NJL model (solid lines), pure hadron matter by BHF 
with Bonn B potential (dashed lines) and the extreme cases of mixed phase by bulk Gibbs 
condition (dotted lines) and Maxwell construction (bold solid lines). 
Each dot shows the mixed phase considering {\it finite size effects},  where inhomogeneous 
structures, such as droplet, rod, slab, tube~(anti-rod), bubble~(anti-bubble), appear. 
The mixed phase appears from the onset density 0.67~(0.70) fm$^{-3}$ to 1.00~(0.97) fm$^{-3}$ 
for $\sigma=10~(40)$ MeV fm$^{-2}$ and $\eta_V=0.20$.
}
\end{figure}

\section{DISCUSSION and future works}
We do not consider the hyperons, since their appearance is likely prevented by that of quarks [15,16]. 
We plan to investigate different quark models and to take into account further degrees of 
freedom in the mixed phase as, e.g., hyperons. 

\acknowledgements{NY likes to express hearty thanks to the organizers on the conference 
``{\it Quarks and compact stars 2014}" for their warm hospitality, DB acknowledges support
by NCN grant UMO-2011/02/A/ST2/00306. This work was supported in part by the COST Action 
MP1304 "NewCompStar".}

\end{document}